\documentclass[aps,amsmath,article,amssymb,amsfonts,twocolumn]{revtex4-2}
\usepackage{graphicx}
\usepackage[colorlinks=true,linkcolor=blue,citecolor=blue, urlcolor=blue]{hyperref}
\usepackage{blindtext}

\usepackage{lineno}

\usepackage{setspace}
\AtBeginDocument
{
	\addtolength\abovedisplayskip{0.20\baselineskip}%
	\addtolength\belowdisplayskip{0.20\baselineskip}%
}

\begin{document}

\title{Fast protein folding is governed by memory-dependent friction}

\author{Benjamin A. Dalton\thanks{coo}}
\affiliation{Freie Universit\"at Berlin, Fachbereich Physik, 14195 Berlin, Germany}
\author{Cihan Ayaz}
\affiliation{Freie Universit\"at Berlin, Fachbereich Physik, 14195 Berlin, Germany}
\author{Lucas Tepper}
\affiliation{Freie Universit\"at Berlin, Fachbereich Physik, 14195 Berlin, Germany}
\author{Roland R. Netz }
\affiliation{Freie Universit\"at Berlin, Fachbereich Physik, 14195 Berlin, Germany}

\begin{abstract}
When described by a low-dimensional reaction coordinate, the rates of protein folding are determined by a subtle interplay between free-energy barriers and friction. While it is commonplace to extract free-energy profiles from molecular trajectories, a direct evaluation of friction is far more elusive, and one typically evaluates it indirectly via memoryless reaction rate theories. Here, using memory-kernel extraction methods founded on a generalised Langevin equation (GLE) formalism, we directly calculate the memory-dependent friction for eight fast-folding proteins, taken from a published set of large-scale molecular dynamics protein simulations. Our results reveal that, contrary to common expectation, friction is more important than free energy barriers in determining protein folding rates, particularly for larger proteins. We also show that proteins fold in a regime where the finite decay time of friction significantly reduces the folding times, in some instances by as much as a factor of 10, compared to predictions based on memoryless friction.
\end{abstract}

\maketitle

\section*{Introduction}

For most proteins, functionality depends on successfully folding into a specific three-dimensional conformational state. This requires that a linear polypeptide chain is driven to explore conformation space by interactions with a solvating environment and is shaped by both solvent interactions and internal interactions between amino acids. When described by a low-dimensional reaction coordinate, folding and unfolding are typically described by a free energy landscape \cite{Bryngelson_1989, Bryngelson_1995, Dill_1997, Schuler_2008,Hinczewski_2013, Plotkin_2003}, with distinct states separated by free energy barriers. For proteins that fold in less than 100 $\mu$s, so-called ``fast-folding" proteins, barrier heights determined from experiment, theory, and simulation are of the order of just a few $k_{\rm{B}}T$'s in height. Theories for describing protein folding are often phrased as reaction rate theories with an explicit dependence on the free energy barrier and friction. An accurate understanding of the kinetics of barrier-crossing reactions is not only important for describing protein folding but is essential in many other fields, such as nucleation theory and chemical kinetics. Reaction rate theory dates back to Arrhenius, who, in 1889, \cite{Arrhenius_1889} showed that the transition times between reactant and product scale according to $\kappa \text{e}^{U_0/k_{\rm{B}}T}$, where $U_0$ is the height of the energy barrier separating the two states. In principle, the pre-factor $\kappa$ can depend on both the free energy profile and friction. Building on the transition state theory of Eyring, \cite{Eyring_1935}, Kramers was the first to include an explicit dependence on solvent friction \cite{Kramers_1940}. The friction in Kramers theory is frequency-independent, suggesting that the environment of the reacting system, and possibly internal processes, relax infinitely fast compared to the barrier crossing time. Regardless of its simplicity, Kramers theory predicts reaction rates in both the over-damped and inertia-dominated regimes well and is sufficient for describing many systems. Various advancements have been made that bridge the over-damped and inertia-dominated regimes, with many accommodating finite relaxation times \cite{Melnikov_1986, Grote_1980, Kappler_2018, Kappler_2019, Lavacchi_2020}. In the case of proteins, it is known that friction is determined by a combination of interactions with the solvent environment and internal intramolecular interactions \cite{Ansari_1992, Echeverria_2014,Schulz_2012,DeSancho_2014}, and that finite relaxation occurs \cite{Lange_2006}. A direct evaluation of the friction acting on a protein has not been possible and approaches have rather relied on determining friction indirectly from memoryless reaction rate theory \cite{Best_2006,Best_2010, Hinczewski_2010, Chung_2015}. In this paper, we directly extract the friction acting on the conformational dynamics of proteins from molecular dynamics simulation trajectories, which accounts for both internal and external friction contributions, and finite relaxation times.

We evaluate friction for eight fast-folding proteins, obtained from previously published large-scale molecular dynamics simulation trajectories \cite{Lindorff_2011}. These cutting-edge simulations, performed by the Shaw group using the purpose-built Anton super-computer \cite{Shaw_2009, Shaw_2010}, represent a breakthrough in the simulation-of-scale, enabling all-atom simulation of fast-folding proteins that would otherwise not be practically possible. These simulations, which were originally performed to reveal the structural and kinetic details of the folding mechanisms for fast-folding proteins, yield simulation trajectories for proteins with lengths that range between 10 and 80 amino-acid residues and simulation times between approximately 100 $\mu$s and 3 ms. These trajectories represent a diverse protein data set, comprised of mixtures of $\alpha$-helix and $\beta$-hairpin secondary structures, and an assortment of tertiary structures, where all proteins execute a multitude of folding and unfolding events.

We analyse all protein trajectories in the framework of a non-Markovian, generalised Langevin equation (GLE). While non-Markovian effects have been investigated in protein folding before \cite{Plotkin_1998,Berezhkovskii_2018} and discrete Markov-state models, which describe folding as transitions between a set of metastable states \cite{Noe_2007, Chodera_2007}, have also been extended to include memory effects \cite{Cao_2020}, this is the first investigation with such extensive simulation data where full GLE memory kernel extraction methods have been applied. 

The GLE is a low-dimensional representation of some higher dimension system. In the present case, we collapse the all-atom dynamics of the composite water-protein systems onto a one-dimensional reaction coordinate. This process of dimension reduction is known as projection \cite{Zwanzig_1961, Mori_1965}. We project the all-atom trajectories, as provided by the Shaw group, onto a well-known fraction of native contacts reaction coordinate, originally introduced by Shakhnovich \textit{et. al.} \cite{Shakhnovich_1991}, with soft cut-off $Q$ \cite{Best_2013}, and hence extract friction memory kernels from $Q(t)$. Free-energy profiles and friction result from dimension reduction, and are therefore unique to a given reaction coordinate. The extracted friction kernels encode dissipation across a spread of time scales, representing the finite relaxation rates of solvent processes and internal reconfigurations, as far as they are represented in $Q(t)$. Consequently, the friction kernel contains the information for the full friction acting on the reaction coordinate. From the free energy landscape and friction kernels, we derive predictions for the folding times on different levels of reaction rate theory and compare to the folding times measured in the simulation. In doing so, we show that the best prediction for the measured folding times is given by a multi-time-scale, non-Markovian theory, which has hitherto only been applied to the characterisation of model systems \cite{Kappler_2019, Lavacchi_2020}. This validation is only possible owing to the long simulation times and diverse range of proteins that comprise the extensive protein data set. We show that the memory decay-times, i.e., the duration of memory effects, are significantly long for all proteins in this set, in some instances as long as the folding times. This indicates that even for $Q$, which is typically considered by other measures to be a good reaction coordinate \cite{Best_2013}, must be considered as a poor reaction coordinate, when judged according to its non-Markovianity. Finally, we also show that, for this particular set of proteins, friction is more important than free-energy barriers in determining folding times, and that this dominance of friction increases for larger proteins. Taken together, our findings suggest that, when represented by a low-dimensional reaction coordinate, the fast conformational dynamics of proteins is dominated by friction and is non-Markovian in nature.

\section*{Results}

\begin{figure*}[t!]
\includegraphics[scale=0.88]{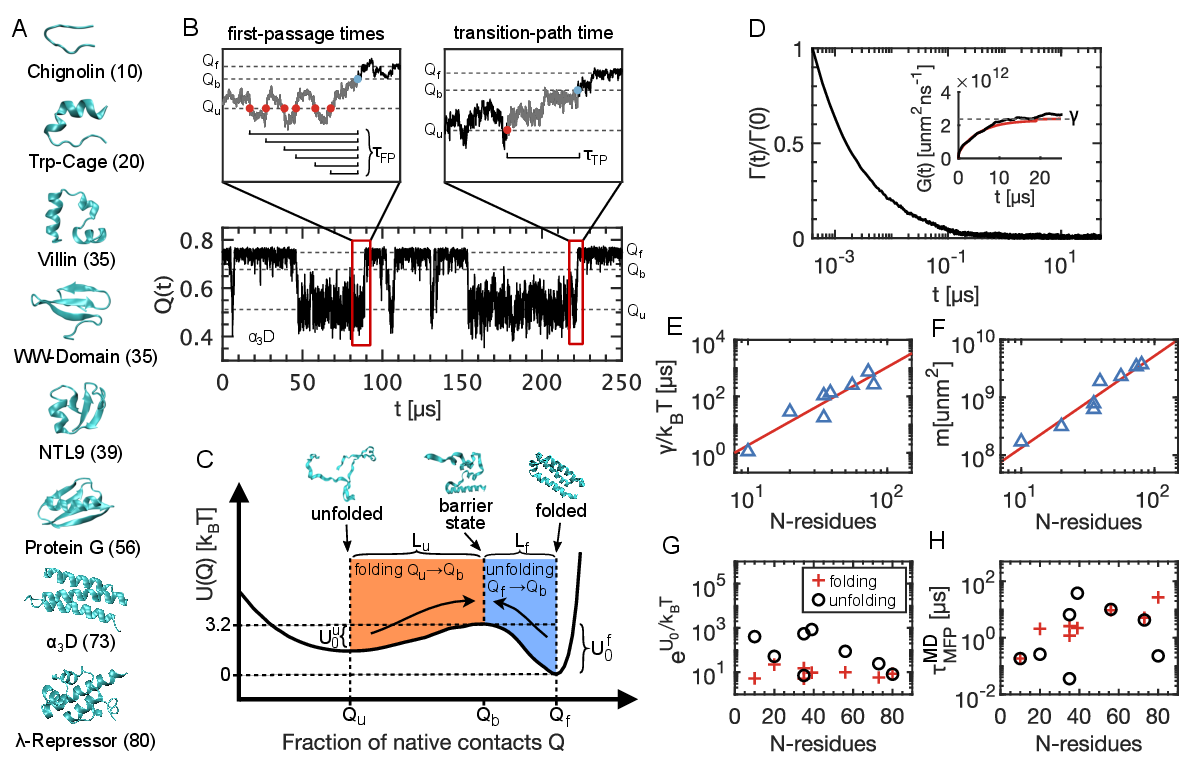}
\caption{The folding and unfolding of eight fast-folding proteins. A) Native states for eight proteins, with the number of amino acids in each protein. B) 250 $\mu$s trajectory segment for the $Q(t)$ reaction coordinate ($\alpha_3$D protein). Left magnification: a sequence of folding first-passage events, from the unfolded state $Q_{\rm{u}}$ to the barrier top $Q_{\rm{b}}$. Right magnification: an example folding transition path and corresponding transition path time $\tau_{\text{TP}}$. C) Free energy profile for the $\alpha_3$D protein. Configuration snapshots show example folded and unfolded states, with an example barrier state. $Q_{\rm{u}}$ and $Q_{\rm{b}}$ are the reaction coordinate values in the folded and unfolded states, respectively. $Q_{\rm{b}}$ is the transition barrier top. Asymmetric barrier heights are also indicated. For $\alpha_3$D, the barrier faced by the unfolded protein, $U^{\rm{u}}_{0} = U(Q_{\rm{b}})-U(Q_{\rm{u}}) = 1.8 \;k_{\text{B}}T$, is less than the barrier faced by the folded protein $U^{\rm{f}}_{0} = U(Q_{\rm{b}})-U(Q_{\rm{f}}) = 3.2 \;k_{\text{B}}T$. The distance from the unfolded state to the barrier top $L_{\rm{u}} = Q_{\rm{b}}-Q_{\rm{u}}$ is greater than the distance from the folded state to the barrier top $L_{\rm{f}} = Q_{\rm{f}}-Q_{\rm{b}}$. D) Normalised GLE memory kernel $\Gamma(t)$, extracted from $Q(t)$ for the $\alpha_3$D protein. Running integral $G(t)$ (black line), the limiting total friction $\gamma$ (inset - dashed line), and an exponential fit (red curve). E) Total friction $\gamma$ for each protein, plotted as a function of the number of residues $N$, divided by $k_{\text{B}}T$ (power law with exponent $\sim2.8$ (red line)). F) Effective mass $m$, plotted as a function of $N$ (power law with exponent $\sim1.5$ (red line)). G) and H) show the Arrhenius factor $\text{e}^{\text{U}_0/k_{\text{B}}T}$ and barrier crossing times $\tau_{\rm{MFP}}^{\text{MD}}$ for folding and unfolding transitions individually. \label{Fig_1}}
\end{figure*}

\begin{figure*}[t!]
\includegraphics[scale=0.77]{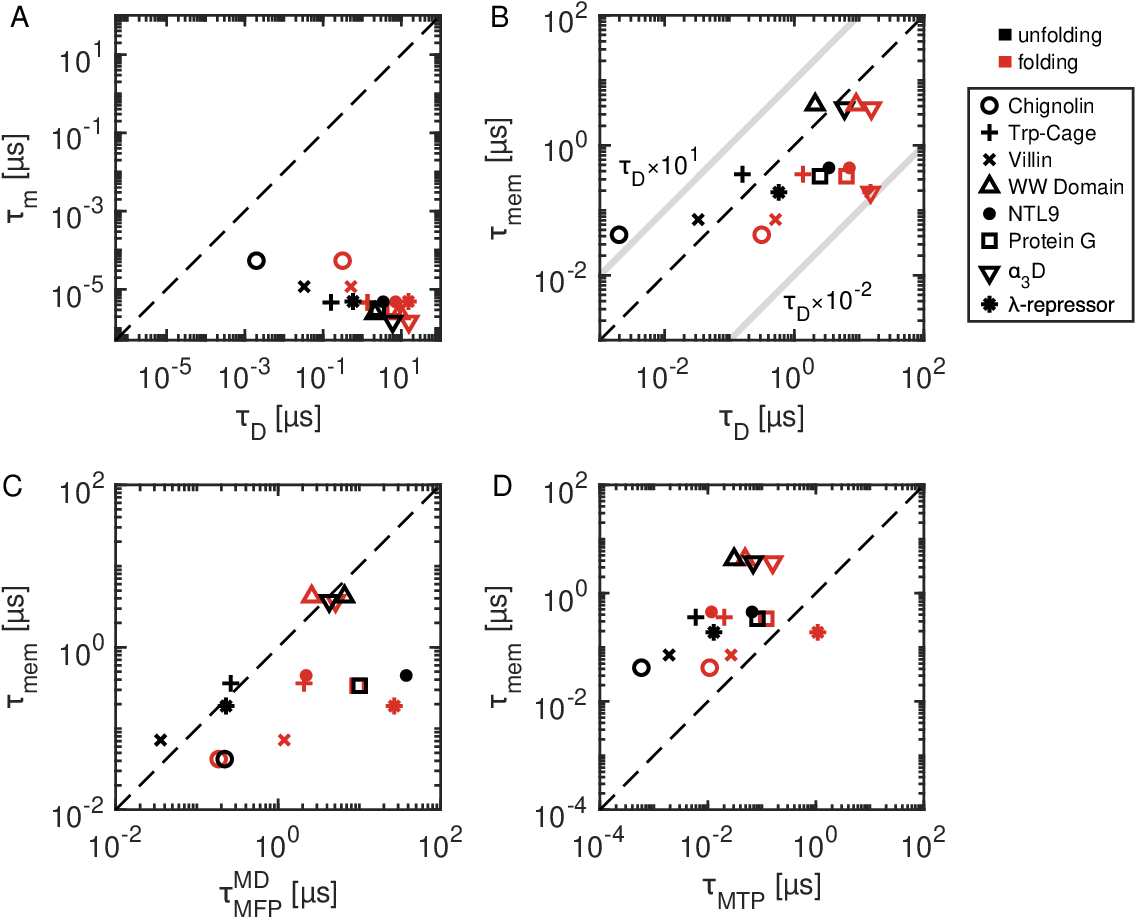}
\caption{Comparison of relevant time scales for protein folding and unfolding kinetics. A) The inertia time scales $\tau_{\rm{m}}$ compared to the diffusion times $\tau_{\rm{D}}$, showing that all systems are in the over-damped regime. B) Memory decay time scales $\tau_{\rm{mem}}$, calculated from the first moment of the memory kernel $\Gamma(t)$, compared to the diffusion times $\tau_{\rm{D}}$ . The light grey lines indicate the bounding regime for $\tau_{\rm{mem}}$ between $\tau_{\rm{D}}{\times}10^{-2} $ and $\tau_{\rm{D}}{\times}10^{1}$, which is the domain in which memory-induced kinetic acceleration is expected. C) Memory times $\tau_{\rm{mem}}$ compared to the folding and unfolding times, expressed as the mean first passage times  $\tau_{\rm{MFP}}^{\text{MD}}$. D) Memory times compared to the transition path times  $\tau_{\rm{MTP}}$ leading from the folded and unfolded state minima to the barrier tops. The broken lines in each plot indicate exact equivalence between the respective times. \label{Fig_2}}
\end{figure*}

\noindent \textbf{Extracting friction from protein simulations:} The native state structures that we determine for the eight proteins, along with the number of residues that make up each chain, are shown in Fig.~\ref{Fig_1}A. The fraction of native contacts reaction coordinate $Q$ is described in detail in the Materials and Methods section. In brief, $Q$ provides a contact-based measure for the deviation of a given state away from the native configuration, which typically represents the folded state. $Q$, which is evaluated in connectivity space and is therefore unitless, is closer to $Q=1$ in the folded state and closer to $Q=0$ when unfolded. In Fig.~\ref{Fig_1}B, we show a 250 $\mu$s trajectory segment for the $\alpha_3$D protein (for a summary of simulation details, see Supplementary Information Section 1). The distinct folded and unfolded states are discernible, located at $Q_{\rm{f}}=0.75$ and $Q_{\rm{u}}=0.51$ respectively, separated by a barrier at $Q_{\rm{b}}=0.66$. Throughout this paper, we consider folding transitions as leading from the unfolded state minimum to the barrier top ($Q_{\rm{u}}\rightarrow Q_{\rm{b}}$), and unfolding transitions as leading from the folded state minimum to the barrier top ($Q_{\rm{f}}\rightarrow Q_{\rm{b}}$). This accounts for the strong asymmetries observed in the free energy profiles for the eight proteins. In the left magnification above the trajectory, we identify a sequence of folding first-passage events. The mean of all such first-passage events, taken over the full trajectory, give the mean first-passage time $\tau_{\rm{MFP}}^{\text{MD}}$ for the folding reaction. Likewise, we evaluate the unfolding times as the mean of all first-passage events from $Q_{\rm{f}}$ to $Q_{\rm{b}}$. In the right magnification, we show an example folding transition path, connecting the unfolded state minimum to the barrier top. A mean transition path time $\tau_{\rm{MTP}}$ is simply the mean for all such transition path durations for either folding or unfolding transition. From the trajectories of $Q(t)$ for each protein, we calculate the free energy profile $U(Q)$ acting on the reaction coordinate. In Fig.~\ref{Fig_1}C, we show $U(Q)=-k_{\rm{B}}T\text{log}[\rho(Q)]$ for the $\alpha_3$D protein, where $\rho(Q)$ is the probability density for $Q(t)$, $k_{\rm{B}}$ is Boltzmann's constant, and $T$ is the system temperature, which has a unique value for each protein. We show the free energy profiles for all proteins in the Supplementary Information Section 2. All free energy profiles are asymmetric, such that the barrier heights faced by the folded and unfolded states ($U^{\rm{u}}_{0}$ and $U^{\rm{f}}_{0}$) are not equal. Likewise, the distances in reaction-coordinate space from the minima to the barrier top ($L_{\rm{f}}$ and $L_{\rm{u}}$) are also different.

Central to this paper is the extraction of time-dependent friction kernels $\Gamma(t)$ from the $Q(t)$ trajectories for each of the eight proteins. We describe the stochastic dynamics of $Q(t)$ by the one-dimensional generalised Langevin equation (GLE)
\begin{equation}\label{GLE_intro}
\begin{split}
m\ddot{Q}(t) &= -\int\limits_{0}^{t}\Gamma(t-t^{\prime})\dot{Q}(t^{\prime})dt^{\prime} \\
& \quad\quad\quad\qquad - \nabla U\big[Q(t)\big]+  F_R(t),
\end{split}
\end{equation}
where $\Gamma(t)$ is the friction memory kernel, $F_R(t)$ is the random force term satisfying the fluctuation-dissipation theorem $\langle F_R(t) F_R(t^{\prime}) \rangle =k_{\rm{B}}T\Gamma(t - t^{\prime})$, $m$ is the effective mass of the reaction coordinate, and $U(Q)$ is the free energy profile. Eq.~\ref{GLE_intro} neglects non-linear friction and is therefore approximate, but has been shown to be accurate for peptide \cite{Ayaz_2022} and chemical bond dynamics \cite{Brunig_2022}. There are various methods for extracting dynamic friction from discrete time-series trajectories \cite{Straub_1987, Daldrop_2017, Daldrop_2018, Kowalik_2019, Darve_2009}. We describe the method used for extracting $\Gamma(t)$ from $Q(t)$ in detail in the Supplementary Information Section 3. In short, we directly extract the running integral of the memory kernel $G(t)=\int_0^t \Gamma(t^{\prime})dt^{\prime}$ using a Volterra extraction scheme, which is suitable for arbitrary, non-linear, free-energy profiles $U(Q)$ \cite{Ayaz_2021, Kowalik_2019}. The memory kernel is then given by $\Gamma(t) = dG(t)/dt$, which is evaluated numerically. The memory kernel for the $\alpha_3$D protein, normalised by $\Gamma(0)$, is shown in Fig.~\ref{Fig_1}D. The dynamics of the $\alpha_3$D protein exhibit significant memory effects, which is clear from the inset, which shows that the running integral $G(t)$ plateaus only after about 10 $\mu$s. We show the memory kernels for all proteins in the Supplementary Information Section 4. \\

\noindent \textbf{Friction is more important than free energy barrier heights:} Having extracted the full time-dependent friction kernel via $G(t)$, we also have access to the zero-frequency friction $\gamma = G(t \rightarrow \infty)$, which we refer to throughout as the total friction. In Fig.~\ref{Fig_1}E, we show $\gamma$ for each protein, plotted as a function of the number of residues in each chain $N$. Since each system has a unique temperature, we divide by $k_{\text{B}}T$. We fit a power law and find that $\gamma/k_{\text{B}}T=3.2{\times}10^{-3}N^{2.8}\;\mu\text{s}$ (red line). For a normalised reaction coordinate that is a sum of $N$ uncorrelated atomic distances, one expects linear scaling in $N$ (see Supplementary Information Section 5). $Q$ is a highly non-linear reaction coordinate. Plus, it is known that reptation dominates the dynamics of collapsed polymers \cite{DeGennes_1971}. Both of these effects contribute to the super-linearity in $N$-scaling for total friction. We also see that the effective mass of $Q$, which we calculate using the equipartition theorem $m=2k_{\text{B}}T/\langle \dot{Q} \rangle^2 $, clearly depends on $N$ (Fig.~\ref{Fig_1}F). The power law scaling is given by $m=3.4{\times}10^{6}N^{1.5}\;\text{unm}^2$ (red line). The super-linearity here is solely due to the non-linearity of $Q$ \cite{Ayaz_2022}. Interestingly, increasing the chain length does not appear to directly couple to free energy barrier heights. The Arrhenius barrier terms $\text{e}^{\text{U}_0/k_{\text{B}}T}$, plotted in Fig.~\ref{Fig_1}G, show that the hindrance to protein folding imposed by free energy barriers is remarkably uncorrelated with the size of a protein. The folding and unfolding kinetics, however, which we quantify as the mean first-passage times $\tau_{\rm{MFP}}^{\rm{MD}}$ (Fig.~\ref{Fig_1}H), do increase with $N$. Therefore, whereas the folding and unfolding barrier heights do not appear to increase with $N$, the barrier crossing times and friction do. Overall, Fig.~\ref{Fig_1}E-H tells us that, when projected onto $Q$, friction plays a dominant role in governing the dynamics and kinetics of protein folding and that the free energy profiles, while certainly also essential, appear to be less influential than friction in determining protein folding reaction rates. This statement will be made more quantitative below. \\

\begin{figure*}[t!]
\includegraphics[scale=0.77]{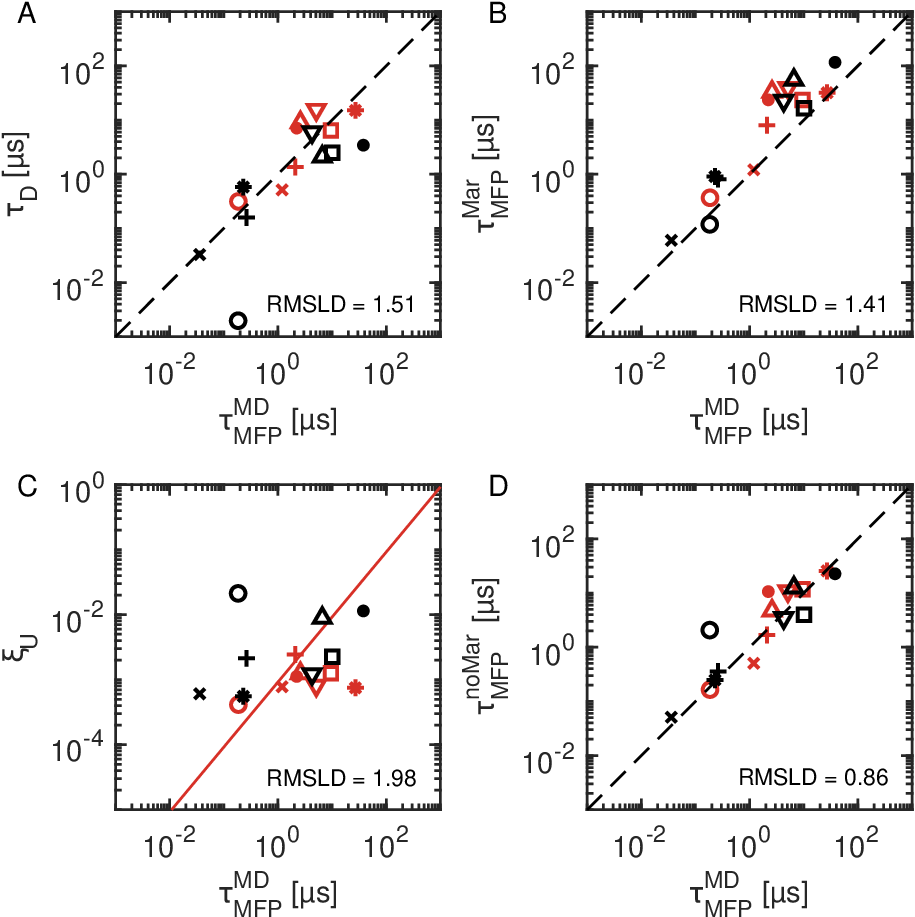}
\caption{Comparison of simulated   protein  folding and unfolding times $\tau_{\rm{MFP}}^{\rm{MD}}$  with predictions  on different levels of theory. 
A) Diffusion times $\tau_{\rm{D}}$, according to Eq.~\ref{tau_D}.
B) Markovian predictions $\tau_{\rm{MFP}}^{\rm{Mar}}$ using the free-energy profiles, according to Eq.~\ref{MPT_Exact}. 
C) Free-energy dependent factor $\xi_{\text{U}}$ that accounts for the effects of the free-energy landscape but not for friction, 
 according to Eq.~\ref{MPT_Exact}. 
D) Non-Markovian predictions $\tau_{\rm{MFP}}^{\rm{noMar}}$, according to Eq.~\ref{tau_noMar}.
 Root-mean-squared logarithmic deviations (RMSLD)   score the population deviation from $\tau_{\rm{MFP}}^{\rm{MD}}$ for A), B), and D) (dashed lines), 
 and from the optimal linear regression in C) (red line). See Fig.~\ref{Fig_2} for symbol legend. \label{Fig_3}} 
\end{figure*}

\noindent \textbf{Memory duration is significant in protein folding:} As well as providing the total friction $\gamma$, the extraction of $\Gamma(t)$ provides a time scale for the sustain of memory effects. Furthermore, $\gamma$ can be used predictively to evaluate other key dynamic time scales, which we now compare. The inertia time $\tau_{\text{m}}=m/\gamma$ is the time scale beyond which the system becomes diffusive \cite{Zwanzig_Book}. The diffusion time,
\begin{equation}\label{tau_D}
\tau_{\rm{D}}=\frac{\gamma L^2}{k_BT},
\end{equation}
is the time taken for a Brownian particle to diffuse over a characteristic distance $L$ in the absence of free energy gradients. Both $\tau_{\text{m}}$ and $\tau_{\rm{D}}$ depend on $\gamma$. A system is in the over-damped regime when $\tau_{\text{m}} \ll \tau_{\rm{D}}$. This condition is met for all eight proteins, for both $L_{\rm{u}}$ and $L_{\rm{f}}$ (Fig.~\ref{Fig_2}A). One ambiguous case is Chignolin unfolding, for which $\tau_{\rm{m}}/\tau_{\rm{D}}=0.027$. The reason is that for Chignolin, both the characteristic length in the folded state, $L_{\rm{f}}$, and $\gamma$ are considerably small. 

By comparing $\tau_{\rm{D}}$ to the decay times of the extracted memory kernels, we can assess whether we expect memory effects to influence barrier-crossing processes \cite{Kappler_2018}. Memory-dependent friction typically exhibits cascading time scales, meaning that $\Gamma(t)$ is multi-modal, with time scales spanning many orders of magnitude \cite{Ayaz_2021}. To assign a single effective time-scale to the memory decay, which we call $\tau_{\rm{mem}}$, we evaluate the first-moment for the set of extracted memory kernels: $\tau_{\rm{mem}} = \int_{0}^{\infty} t \Gamma(t)dt/\int_{0}^{\infty} \Gamma(t)dt$. In Supplementary Information Section 6, we address the issue of discretisation due to the intrinsically low time resolution of the MD data with a highly resolved trajectory for an $\alpha$-helix-forming Ala$_9$ homo-peptide chain. Fig.~\ref{Fig_2}B shows that, overall, memory times are non-negligible when compared to diffusion times. It has been shown that memory accelerates barrier crossing kinetics in the memory-time range $1{\times}10^{-2} < \tau_{\rm{mem}}/\tau_{\rm{D}} <1{\times}10^{1}$ \cite{Kappler_2018, Kappler_2019, Lavacchi_2020} (indicated in Fig.~\ref{Fig_2}B), where the Grote-Hynes theory is valid \cite{Grote_1980}. Thus, for the proteins considered here, memory is predicted to accelerate protein folding and unfolding, which we quantify further below. In Fig.~\ref{Fig_2}C, we see that memory times are at most comparable to the reaction times but are typically shorter, indicating that $Q$ is a poor reaction coordinate. The overall concurrence between $\tau_{\rm{mem}}$, $\tau_{\rm{D}}$, and $\tau_{\rm{MFP}}^{\text{MD}}$ predicts a coupling between memory effects and folding kinetics.

The comparison between memory times and transition path times is particularly revealing. Transition paths are the segments of a trajectory where the protein actually executes the reconfiguration from one specific state to another target state. Transition paths are of much interest since they provide the actual folding mechanisms of a protein, and have been studied extensively in experiments \cite{Chung_2013, Chung_2012, Chung_2015, Neupane_2016,Neupane_2016B,Sturzenegger_2018,Zijlstra_2020}, and in the context of simulation and theory \cite{Lindorff_2011, Best_2013, Satija_2017, Daldrop_2016,Hummer_2003,Laleman_2017, Carlon_2018}. In Fig.~\ref{Fig_2}D, we show that for all proteins, transition path times are short compared to memory times. This is significant for two reasons. Firstly, it tells us that entire transitions either fold or unfold under the influence of memory effects that linger from the preceding state. Secondly, it explains why the approximate, position-independent form of Eq.~\ref{GLE_intro} works. Even describing the dynamics of $Q$ as having distinct memory dependence on separate sides of the barrier would be neutralised since memory effects can last for as much as two orders of magnitude longer than the time required to transition between states. \\

\begin{figure}[t!]
\includegraphics[scale=0.82]{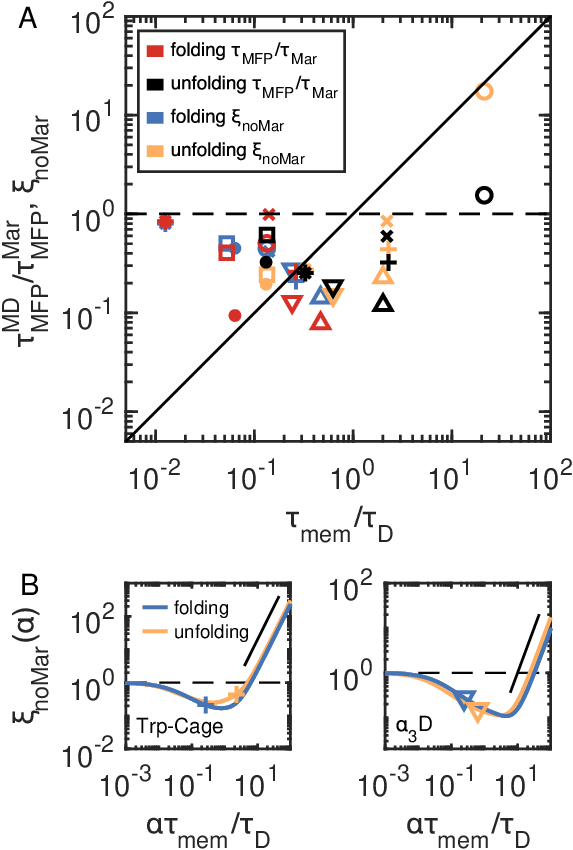}
\caption{Barrier crossing times indicate memory-induced speed up. A) Deviations from Markovian barrier crossing kinetics, plotted as a function of scaled memory times $\tau_{\rm{mem}}/\tau_{\rm{D}}$, for the folding and unfolding of all eight proteins. Rescaled MD simulation values for each protein ($\tau_{\rm{MFP}}^{\text{MD}}/\tau_{\rm{MFP}}^{\rm{Mar}}$, red and black symbols) are compared to multi-modal, non-Markovian prediction ($\xi_{\text{noMar}}$ Eq.~\ref{xi_main}, blue and yellow symbols). See Fig.~\ref{Fig_2} for symbol legend. B) Multi-modal, non-Markovian predictions $\xi_{\text{noMar}}$ plotted as a function of rescaled first-moment memory time $\alpha\tau_{\rm{mem}}$ for two examples: Trp-Cage and $\alpha_3$D. The symbols are for $\alpha=1$, which coincide with the appropriate $\xi_{\text{noMar}}$ values in A). Black lines indicate quadratic scaling $\sim\big(\tau_{\rm{mem}}/\tau_{\rm{D}}\big)^2$ for the long memory-time slow-down regime. Dashed black lines show unity throughout. \label{Fig_4}}
\end{figure}

\noindent  \textbf{Non-Markovian rate theory predicts protein folding times:} 
From the simulation trajectories, we extract  the folding and unfolding times, evaluated via the mean first-passage times $\tau_{\rm{MFP}}^{\text{MD}}$. Since we also extract the friction directly from the simulation trajectories, we can compare the performance of various theoretical estimates for folding and unfolding times. To quantify the population-wide deviation away from $\tau_{\rm{MFP}}^{\text{MD}}$, we use the root-mean-square logarithmic deviation (RMSLD) (see Supplementary Information Section 7). This is defined by $\text{RMSLD}=\sqrt{\sum_{i=1}^n(\text{log}(\tau^{\text{MD}}_{\text{MFP},i}) - \text{log}(\tau^{\text{theo}}_{\text{MFP},i}))^2/n}$, where $\tau^{\text{MD}}_{\text{MFP},i}$ is the value from MD, $\tau^{\text{theo}}_{\text{MFP},i}$ is the theoretical predicted value for a given method, and $n=16$ is the total number of folding and unfolding reactions. Using the diffusion times $\tau_{\rm{D}}$ given in Eq.~\ref{tau_D} as the simplest predictor, we determine a total logarithmic deviation of 1.51, see Fig.~\ref{Fig_3}A for the correlation plot. This agreement is remarkable, since it completely neglects the free energy barriers and depends only on friction and the characteristic separations in Q space, which implies that friction alone provides a good approximation for protein folding times. The Markovian prediction, which in addition to friction explicitly accounts for the extracted free energy profile \cite{Zwanzig_Book}, is given by 
\begin{equation}\label{MPT_Exact} 
\begin{split}
&\tau_{\rm{MFP}}^{\rm{Mar}}\big(Q_\text{s}|Q_\text{e}\big)  =  \tau_{\text{D}}\xi_{\text{U}} \\
& = \tau_{\rm{D}}\int_{Q_\text{s}}^{Q_\text{e}}\text{e}^{ U(x)/(k_{\rm{B}}T) }
\bigg[\int_{-\infty}^{x}\text{e}^{- U(y)/(k_{\rm{B}}T)}\frac{dy}{L}\bigg]\frac{dx}{L}.
\end{split}
\end{equation}
Eq.~\ref{MPT_Exact} assumes constant friction across $Q$ and evaluates the mean first-passage times between a start and end-point on the free energy landscape, $Q_\text{s}$ and $Q_\text{e}$. For folding reactions, for example, we calculate $\tau_{\rm{MFP}}^{\rm{Mar}}(Q_{\text{u}}|Q_{\text{b}})$ (see Supplementary Information Section 8 for details and the expression for unfolding reactions). Again, we see that the population-wide prediction is good (Fig~\ref{Fig_3}B), but with a general trend that the Markovian predictions are slow compared to the MD data. The RMSLD value of 1.41 for the Markovian predictions shows that the improvement caused by multiplying the diffusion time $\tau_{\rm{D}}$ by the free-energy dependent factor  $\xi_{\text{U}}$, defined in Eq.~\ref{MPT_Exact}, is surprisingly small. To emphasise  this, in Fig~\ref{Fig_3}C we correlate  $\xi_{\text{U}}$ with the MD barrier crossing times and obtain a RMSLD value of 1.98 using linear regression (red line), a value that is significantly higher than that obtained for the diffusion times in Fig~\ref{Fig_3}A. In other words, friction describes the variation of folding and unfolding times among the considered set of proteins more accurately than the free energy barriers. It transpires that the prefactor of the Arrhenius exponential factor is more important than the exponential itself.

Can we improve  the Markovian prediction  in  Eq.~\ref{MPT_Exact}? We define  a non-Markovian correction factor $\xi_{\text{noMar}}$ as 
\begin{equation}   \label{xi_main}
\xi_{\text{noMar}} = \frac{\tau^{\text{H}}_{\rm{MFP}}(\tau_{\rm{D}},U_0, \lbrace \gamma_i \rbrace,\lbrace \tau_i \rbrace,\tau_{\rm{m}}   )}{\tau_{\rm{MFP}}^{\text{H}}(\tau_{\rm{D}},U_0, \lbrace\gamma_i \rbrace, 0 ,0)}.
\end{equation}
Here, $\tau^\text{H}_{\rm{MFP}}$ is  a recently proposed closed-form heuristic expression for the barrier-crossing time of a one-dimensional reaction coordinate characterized by a finite mass  in the presence of multi-exponential memory  $\Gamma(t) = \sum_{i=1}^M (\gamma_i/\tau_i)\exp{(t/\tau_{i})}$, characterised by  $M$ friction factors $\lbrace \gamma_i \rbrace$ and $M$  memory times $\lbrace \tau_i \rbrace$ \cite{Kappler_2019, Lavacchi_2020} (see materials and methods section Eqs.~\ref{OD} - \ref{Heurist} for details of $\xi_{\text{noMar}}$ and $\tau^\text{H}_{\rm{MFP}}$ ). The multi-exponential fits of our extracted memory kernels are explained in  Supplementary Information Section 4, for Chignolin we use  $M$=2 and for all other proteins $M$=3. The denominator in Eq.~\ref{xi_main} represents the overdamped Markovian limit, where the mass and all memory times are set to  zero.  The non-Markovian prediction for the barrier-crossing time can thus be written as
\begin{equation}\label{tau_noMar}
\tau_{\rm{MFP}}^{\text{noMar}}=\xi_{\text{noMar}} \tau_{\rm{MFP}}^{\rm{Mar}} = \xi_{\text{noMar}}  \xi_{\text{U}}   \tau_{\rm{D}}.
\end{equation}
In Fig~\ref{Fig_3}D, we see that $\tau_{\rm{MFP}}^{\text{noMar}}$ improves the overall prediction of the MD folding and unfolding times (except for Chignolin, Villin  and Protein G), as affirmed by a RMSLD value of 0.86. While the errors are still quite large, we recall that all parameters needed for the evaluation of $\tau_{\rm{MFP}}^{\text{noMar}}$ are extracted from the MD trajectories, without any fit to the MD reaction times. \\

\noindent \textbf{Proteins fold in the memory-induced speed-up regime:} Memory effects can either speed-up or slow-down barrier-crossing times when compared to the memoryless limit, depending on the ratio of memory and diffusion times $\tau_{\rm{mem}}/\tau_{\rm{D}}$ \cite{Kappler_2018, Kappler_2019, Lavacchi_2020}. According to  our analysis, $\tau_{\rm{MFP}}^{\rm{Mar}}$ as given by Eq.~\ref{MPT_Exact} represents the memoryless  limit. By plotting $\tau_{\rm{MFP}}^{\text{MD}}/\tau_{\rm{MFP}}^{\rm{Mar}}$  against the rescaled memory times $\tau_{\rm{mem}}/\tau_{\rm{D}}$ for all proteins, we quantify the deviation from Markovian behaviour over the range of extracted memory times. The red (folding) and black (unfolding) symbols in Fig.~\ref{Fig_4}A reveal reaction speed-up across the entire population of proteins for intermediate values of $\tau_{\rm{mem}}/\tau_{\rm{D}}$, while in the short memory-time limit $\tau_{\rm{mem}}/\tau_{\rm{D}}\rightarrow 0$, the reaction times trend to Markovian behaviour, i.e. $\tau_{\rm{MFP}}^{\text{MD}}/\tau_{\rm{MFP}}^{\rm{Mar}} \rightarrow 1$. This is precisely what is expected based on previous works on model systems \cite{Kappler_2019, Lavacchi_2020} and on simulations of short homo-peptide chains \cite{Ayaz_2021}. The scaling plot Fig.~\ref{Fig_4}A thus indicates that folding and unfolding times are significantly accelerated by memory effects. In some instances, the folding and unfolding times are accelerated by as much as a factor of 10, which is a remarkable contribution.

This notion can be made more quantitative by using $\xi_{\text{noMar}}$, which, according to Eq.~\ref{tau_noMar}, describes the ratio $\tau_{\rm{MFP}}^{\text{MD}}/\tau_{\rm{MFP}}^{\rm{Mar}}$, predictively. Except for the case of Chignolin unfolding, $\xi_{\text{noMar}}$, as given by  Eq.~\ref{xi_main} and shown as blue (folding) and yellow (unfolding) symbols in Fig.~\ref{Fig_4}A, agrees excellently with the simulation data, indicating that multi-modal non-Markovian reaction rate theory describes well the accelerated barrier-crossing observed in protein folding simulations, even predicting the 10-fold acceleration for intermediate memory times.
  
To gain deeper understanding of how  memory affects protein reaction times, 
we  uniformly rescale all memory times  that enter Eq.~\ref{xi_main} by a factor $\alpha$. Such a rescaling can be interpreted as a variation of the  solvent viscosity. In doing so, we effectively generate an $\alpha$-dependent memory kernel $\Gamma^{\alpha}(t) = \sum_{i=1}^M (\gamma_i/\alpha\tau_i)\exp{(t/\alpha\tau_{i})}$ and a corresponding rescaled first-moment memory  time  $\alpha\tau_{\rm{mem}}=\int_0^{\infty}t\Gamma^{\alpha}(t)dt/\int_0^{\infty}\Gamma^{\alpha}(t)dt$. In Fig.~\ref{Fig_4}B, we show two examples of $\xi_{\text{noMar}}(\alpha)$, plotted as a function of $\alpha\tau_{\rm{mem}}/\tau_{\rm{D}}$. The example of Trp-Cage is particularly interesting, as it shows that the prediction of $\xi_{\text{noMar}}$ for unfolding, as given in Fig.~\ref{Fig_4}A and replotted as the yellow plus symbol  in Fig.~\ref{Fig_4}B, is actually located close to the transition from the speed-up regime, approximately obtained for  $\alpha\tau_{\rm{mem}}/\tau_{\rm{D}} < 10$, to the slow-down regime for  $\alpha\tau_{\rm{mem}}/\tau_{\rm{D}} >10$. This is also true of Villin and the WW-Domain (see Supplementary Information Section 9). In fact, Grote-Hynes theory correctly predicts the memory-induced  reaction speed-up \cite{Grote_1980}, but  misses the memory-induced reaction slow-down that occurs for long memory times and  is characterised by a quadratic scaling of the reaction time with the memory time \cite{Kappler_2018}, which is also displayed for large $\alpha$ in Fig.~\ref{Fig_4}B, as indicated by black straight lines. Overall, the comparison between $\xi_{\text{noMar}}$  for the actual MD parameters and the scaling function $\xi_{\text{noMar}}(\alpha)$ in Fig.~\ref{Fig_4}B shows that proteins fold and unfold in the memory speed-up regime close to the border to the slow-down regime; this reasserts that  non-Markovian effects are crucial in order to quantitatively predict protein reaction rates. \\

\section*{\small Discussion}

We have extracted memory kernels from published large-scale MD simulation trajectories of eight fast-folding proteins \cite{Lindorff_2011} and found that, when measured in the $Q$ reaction coordinate, memory times are comparable to the protein folding and unfolding reaction times and typically vastly exceed the transition path times. There are different definitions of a good reaction coordinate \cite{Plotkin_1998, Sali_1994, Du_1998} and such a quality is often context-dependent. For example, a reaction coordinate might be optimised such that barrier-crossing states are most likely to be transition states \cite{Best_2005}, which has been shown to be the case for $Q$ \cite{Best_2013}. But such definitions do not say anything about whether the dynamics of such a reaction coordinate is Markovian, which we have shown here is certainly not the case for $Q$. In the Supplementary Information Section 10, we show examples of $\Gamma(t)$ for other standard reaction coordinates, such as the end-to-end distance, radius of gyration, and the RMSD from the native state. The resultant memory times span orders of magnitudes but are typical of the order of the diffusion time scale $\tau_{\text{D}}$. Therefore, none of these reaction coordinates can be considered a good reaction coordinate, insofar as optimising for Markovianity is a viable measure of goodness. However, as we  demonstrate in this paper, it does not matter whether a reaction coordinate displays non-Markovian dynamics or not: As long as one uses the appropriate non-Markovian framework for analysing protein folding trajectories, taken from either simulation or experiments \cite{Chung_2013, Chung_2012, Neupane_2016,Pirchi_2016, Aviram_2018}, one can accurately predict the folding kinetics of a protein. In fact, we show that multi-exponential non-Markovian reaction rate theory reliably  predicts  folding and unfolding  times from  MD simulations.

Our results indicate that, for the set of proteins considered in this paper, friction is more important than free energy barriers in determining folding times. The disclaimer is important here, since the protein trajectories that we have analysed come from a biased set of proteins selected for short folding times. Also, simulation temperatures are chosen to maximise  the frequency of folding/unfolding transitions and are close to  the melting temperatures. Our results show that the total friction $\gamma$ acting on $Q$ increases more than quadratically with chain length. Barrier heights do not increase with chain length, but the barrier crossing times do (Fig.~\ref{Fig_1}E-H), clearly showing that friction effects are dominant. We cannot suggest that such dependencies hold for proteins with much larger barrier heights or for general temperatures. Regardless, it appears that, for the present data, contributions from the friction-dependent pre-factor in the Arrhenius-type reaction-rate theories for protein folding dominate the exponential term. This is made clear when we compare Figs.~\ref{Fig_3}A - C: A simple  prediction for the barrier crossing times that only depends on friction, i.e. $\tau_{\text{D}}$, represents well the simulated barrier crossing times. Explicitly accounting for the exact free energy profiles makes very little improvement on this prediction, which means that friction is dominant.  A far greater improvement is made when we account for  non-Markovian effects, as is seen in Fig.~\ref{Fig_3}D.
 
Whether in simulation or experiment, protein conformation dynamics is typically described by some reduced-dimension collective reaction coordinate. Overall, our results suggest that, irrespective of the choice of the reaction coordinate, non-Markovian effects are present and that these effects must be taken into account when attempting to model observable features of folding proteins. 

In Eq.~\ref{tau_noMar}, we take into account non-Markovian effects, but we assume constant friction. Alternative models neglect non-Markovian effects but include position-dependent friction \cite{Hinczewski_2013, Hummer_2003}. In the Supplementary Information Section 11, we show that for a Markovian model with position-dependent friction, there does not exist a unique friction profile $\gamma(Q)$ for describing both folding and unfolding kinetics. This was already shown to be true for a simple $\alpha$-helix forming homo-alanine chain  \cite{Ayaz_2021}.  From this we conclude that we can not predict protein dynamics consistently using Markovian theory with a unique friction profile $\gamma(Q)$. Likewise, we show in Supplementary Information Section 5 that the increase in total friction as a function of protein chain-length is not a consequence of the normalisation included in the evaluation of $Q$ (see Eq.~\ref{Q_nc})

Our analysis uses the  approximate GLE Eq.~\ref{GLE_intro}  with a position-independent memory kernel. In fact, exact formulations of the GLE that include position-dependent memory kernels and at the same time can be fully parameterised from time series data have been recently proposed \cite{Ayaz_2022} and would allow to account for non-Markovian effects as well as position-dependent friction. Such a treatment is presumably important for proteins that exhibit multiple distinct folding or unfolding pathways, which has been shown to be the case for NTL9 and the WW-Domain \cite{Lindorff_2011}, and is left for future work. 

\renewcommand{\theequation}{M\arabic{equation}} 
\setcounter{equation}{0}
	
\section*{\small Methods}

\noindent In the Supplementary Information document, Section 1, we present details for the molecular dynamics simulations, including various relevant simulation parameters. Additionally, we include a range of measured time-scales and other extracted quantities, and we compare results for our analysis of reaction times to those from Lindorff-Larsen \textit{et. al.} \cite{Lindorff_2011}. \\

\noindent \textbf{The fraction of native contacts:} For each protein, we project the back-bone C$_{\alpha}$ atomic positions from the all-atom trajectories taken from \cite{Lindorff_2011} onto the fraction of native contacts reaction coordinate $Q$, evaluated with a soft cut-off potential \cite{Best_2013}. The evaluation of $Q(t)$ requires a reference state, which we take to be the native state for each trajectory. To evaluate the native state, we follow previous implementations and select from amongst the member states of the trajectories (as opposed to using, for example, the PDB entry). The approach is similar to that used by Lindorff-Larsen \textit{et. al.} \cite{Lindorff_2011} and Best \textit{et. al.} \cite{Best_2013}, which follows from \cite{Daura_1999}. Briefly, we sample a subset of evenly spaced states from the full trajectory. For each pair of states, we calculate the corresponding root-mean-squared deviation (RMSD) between the two states. If the RMSD between two states is less than 0.2 nm, then we place the pair into a list. We assign the state that has the most listed pairs satisfying the RMSD condition as the native state for a given protein. The native states for each protein are displayed in Fig.~\ref{Fig_1}A. Note that for proteins with more than one independent trajectory segments we select a single native state from amongst all trajectory segments, which we then use for all segments. In the native state, we define all C$_{\alpha}$ pairs that are separated by at least 5 residues in the primary sequence and which are separated by less than 0.9 nm in Cartesian distance, as the native contacts. Each protein will have $N_{\rm{nc}}$ native contacts. $\textbf{s}_{ij}^0$ are the separation vectors for all native contact pairs in the native state, which have magnitudes $s_{ij}^0=\sqrt{\textbf{s}_{ij}^0\cdot \textbf{s}_{ij}^0}$. $\textbf{s}_{ij}(t)$ are the separation vectors for all native contact pairs at each time, with magnitudes $s_{ij}(t)=\sqrt{\textbf{s}_{ij}(t)\cdot \textbf{s}_{ij}(t)}$. This gives the fraction of the native contacts that are deemed to be in contact at time t as
\begin{equation}\label{Q_nc}
Q(t) = \frac{1}{{N_{\rm{nc}}}}\sum_{i<j}\frac{1}{1+\text{e}^{\beta(s_{ij}(t) - \gamma s_{ij}^0 )}},
\end{equation}
where the summation indices $i$ and $j$ are only for native contact pairs. Here, we set the parameters such that $\beta=30$ nm$^{-1}$ and $\gamma=1.6$. 

\noindent \textbf{Non-Markovian correction factor $\xi_{\text{noMar}}$:} From Kappler \textit{et. al.} \cite{Kappler_2019} and Lavacchi \textit{et. al.} \cite{Lavacchi_2020}, we describe multi-exponential memory dependent barrier crossing times as a sum of contributions from $M$ over-damped contributions $ \lbrace\tau^{i}_{\rm{OD}} \rbrace$, and $M$ energy-diffusion contributions $ \lbrace\tau^{i}_{\rm{ED}} \rbrace$, where $i=1,2,...,M$. The individual contributions are defined as follows. For the over-damped contributions, we have
\begin{equation}\label{OD}
\begin{split}
\tau^{i}_{\rm{OD}} &= \tau_{\text{D}}\frac{\gamma_i}{\gamma}\frac{\text{e}^{\beta U_0}}{\beta U_0}\\
& \qquad {\times}\Bigg[\frac{\pi}{2\sqrt{2}} \frac{1}{1+10\beta U_0\tau_{i}/\tau_{\text{D}}} + \sqrt{\beta U_0\frac{\tau_{\rm{m}}}{\tau_{\text{D}}}} \Bigg],
\end{split}
\end{equation}\\
and for the energy-diffusion contributions
\begin{equation}\label{ED}
\begin{split}
\tau^{i}_{\rm{ED}} &= \tau_{\text{D}}\frac{\gamma}{\gamma_i}\frac{\text{e}^{\beta U_0}}{\beta U_0} \\
& \qquad {\times}\Bigg[\frac{\tau_{\rm{m}}}{{\tau_{\text{D}}}} + 4\beta U_0\bigg(\frac{\tau_{i}}{\tau_{\text{D}}}\bigg)^2   + \sqrt{\beta U_0\frac{\tau_{\rm{m}}}{\tau_{\text{D}}}} \Bigg].
\end{split}
\end{equation}
Here, $\beta = 1/k_{\rm{B}}T$. We combine Eqs.~\ref{OD} and \ref{ED} such that the predicted mean passage times are given by
\begin{equation}\label{Heurist} 
\begin{split}
&\tau^{\text{H}}_{\rm{MFP}}(\tau_{\rm{D}},U_0, \lbrace \gamma_i \rbrace,\lbrace \tau_i \rbrace,\tau_{\rm{m}}   )  \\
&\qquad\qquad\qquad\qquad = \sum_{i=1}^M  \tau^{i}_{\rm{OD}} + \Bigg[ \sum_{i=1}^M  \frac{1}{\tau^{i}_{\rm{ED}}} \Bigg]^{-1}.
\end{split}
\end{equation}
$ \lbrace \gamma_i \rbrace$ and $ \lbrace \tau_i \rbrace$ are the sets of $M$ amplitudes and time-scales that appear in Eqs.~\ref{OD} and \ref{ED}. 
The total friction $\gamma$ that appears in Eqs.~\ref{OD} and \ref{ED} is accounted for since $\gamma=\sum_{i=1}^M\gamma_i$. The overdamped
 Markovian limit is achieved by setting  all memory time scales and inertial times equal to 0 and is given by 
\begin{equation}
\tau_{\rm{MFP}}^{\text{H}}(\tau_{\rm{D}},U_0, \lbrace \gamma_i \rbrace , 0 ,0)= \tau_{\text{D}}\pi U_0 \text{e}^{\beta U_0}  /2\beta  \sqrt{2},
\end{equation}
leading to the non-Markovian correction factor $\xi_{\text{noMar}}$ and the non-Markovian barrier crossing time $\tau_{\rm{MFP}}^{\rm{noMar}}$, as given by Eq.~\ref{xi_main} and Eq.~\ref{tau_noMar}, respectively. 

\section*{\small Acknowledgements} 

\noindent We are grateful to the high-performance computer services at the Freie University of Berlin. This project was funded by the European Research Council (ERC) Advanced Grant 835117 NoMaMemo.






\bibliography{Bib_file.bib}

\end{document}